\documentclass[aps,prl,final,twocolumn,superscriptaddress,%
floatfix,preprintnumbers]{revtex4}

\usepackage{dcolumn} 
\usepackage[tbtags]{amsmath}

\usepackage{graphicx}
\usepackage{floatflt}

\newcommand{\bc}{\begin{center}}
\newcommand{\ec}{\end{center}}
\def\t0{t\text{$=$}0}
\def\ix0{\xi\text{$=$}0}

\newcommand{\Dlr}{{D^{\hspace{-0.8em}%
      \raisebox{0.8ex}{$\scriptstyle\leftrightarrow$}}}{}}
\newcommand{\Dl}{{D^{\hspace{-0.8em}%
      \raisebox{0.8ex}{$\scriptstyle\leftarrow$}}}{}}
\newcommand{\Dr}{{D^{\hspace{-0.8em}%
      \raisebox{0.8ex}{$\scriptstyle\rightarrow$}}}{}}
\newcommand{\gev}{\operatorname{GeV}}
\newcommand{\mev}{\operatorname{MeV}}
\newcommand{\fm}{\operatorname{fm}}
\newcommand{\ms}{\mskip 1.5mu}

\begin{document}  

\title{The spin structure of the pion}

\preprint{DESY 07-120, Edinburgh 2007/13, LTH 754, TUM/T39-07-12}

\author{D.~Br\"ommel}
  \affiliation{Deutsches Elektronen-Synchrotron DESY, 22603 Hamburg, Germany}
  \affiliation{Institut f\"ur Theoretische Physik, Universit\"at
  Regensburg, 93040 Regensburg, Germany}
\author{M.~Diehl}
  \affiliation{Deutsches Elektronen-Synchrotron DESY, 22603 Hamburg, Germany}
\author{M.~G\"ockeler}
  \affiliation{Institut f\"ur Theoretische Physik, Universit\"at
  Regensburg, 93040 Regensburg, Germany}
\author{Ph.~H\"agler}
  \affiliation{Institut f\"ur Theoretische Physik T39,
   Physik-Department der TU M\"unchen, 85747 Garching, Germany}
   \email{phaegler@ph.tum.de}
\author{R.~Horsley}
  \affiliation{School of Physics, University of Edinburgh, Edinburgh
  EH9 3JZ, UK}
\author{Y.~Nakamura}
  \affiliation{John von Neumann-Institut f\"ur Computing NIC / DESY,
  15738 Zeuthen, Germany}
\author{D.~Pleiter}
  \affiliation{John von Neumann-Institut f\"ur Computing NIC / DESY,
  15738 Zeuthen, Germany}
\author{P.E.L.~Rakow}
\affiliation{
%Theoretical Physics Division, 
Department of Mathematical Sciences, University of Liverpool, Liverpool L69 3BX, UK}
\author{A.~Sch\"afer}
  \affiliation{Institut f\"ur Theoretische Physik, Universit\"at
  Regensburg, 93040 Regensburg, Germany}
\author{G.~Schierholz}
  \affiliation{Deutsches Elektronen-Synchrotron DESY, 22603 Hamburg, Germany}
  \affiliation{John von Neumann-Institut f\"ur Computing NIC / DESY,
  15738 Zeuthen, Germany}
\author{H.~St\"uben}
  \affiliation{Konrad-Zuse-Zentrum f\"ur Informationstechnik Berlin,
  14195 Berlin, Germany}
\author{J.M.~Zanotti}
  \affiliation{School of Physics, University of Edinburgh, Edinburgh
  EH9 3JZ, UK}

\collaboration{QCDSF/UKQCD Collaborations}\noaffiliation

\date{\today}

\begin{abstract} 
  We present the first calculation of the transverse spin structure of
  the pion in lattice QCD. 
  Our simulations are based on two flavors of non-perturbatively improved
  Wilson fermions, with pion masses as low as $400\mev$ in volumes 
  up to $(2.1 \fm)^3$ and lattice spacings below $0.1\fm$.
  We find a characteristic asymmetry in the
  spatial distribution of transversely polarized quarks.  This
  asymmetry is very similar in magnitude to the analogous asymmetry we
  previously obtained for quarks in the nucleon.  Our results support
  the hypothesis that all Boer-Mulders functions are alike.
\end{abstract}

\maketitle

{\em Introduction.}---
Since their discovery in the late 1940s, pions have played a central
role in 
nuclear and particle physics. As
pseudo-Goldstone bosons of spontaneously broken chiral symmetry they
are at the core of the low-energy sector of quantum chromodynamics
(QCD).
Since the pion
has spin zero, its longitudinal spin structure in terms of
quark and gluon degrees of freedom is trivial.  Pion matrix
elements of quark and gluon helicity operators 
vanish due to
parity invariance, 
$\langle \pi(P')| \Sigma^3 |\pi(P)\rangle\! =\! 0$,
where, e.g., for quarks $\Sigma^3\!=\!\overline{q}\gamma^3\gamma_5\ms q$.
An instructive quantity describing the spin structure of hadrons is
the probability density $\rho(x,b_\perp)$ of quarks in impact
parameter space \cite{Burkardt:2000za},
illustrated in Fig.~\ref{FigIll}.
\begin{floatingfigure}[r]{0.2\textwidth}
\bc
\includegraphics[width=0.2\textwidth]{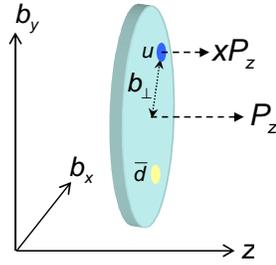}
\caption{\label{FigIll} Illustration of the quark distribution in a
$\pi^+$ in impact parameter space.}
\ec
\vspace{-2mm}
\end{floatingfigure}
\noindent Here $x$ is the longitudinal momentum fraction carried by
the quark, and the impact parameter $b_\perp$ gives the distance
between the quark and the center of momentum of the hadron in the
plane transverse to its motion.
Because 
of parity invariance, the density
$\rho(x,b_\perp,\lambda)$ of quarks with helicity $\lambda$ 
in a pion
is
determined by the unpolarized density, $2\rho(x,b_\perp,\lambda) =
\rho(x,b_\perp)$.  
The latter is given by $\rho(x,b_\perp) = H^\pi(x,\ix0,b_\perp^2)$
in terms of
a $b_\perp$ dependent generalized parton distribution (GPD) at zero skewness $\xi$.
The lattice QCD calculations discussed below give access to
$x$-moments of quark spin densities, which we have investigated in
\cite{Gockeler:2006zu} for  
quarks with transverse spin
$s_\perp$ in a nucleon with transverse spin~$S_\perp$.  The
corresponding expression $\rho(x,b_\perp,s_\perp)$ for polarized
quarks in the pion is  
obtained by setting $S_\perp=0$ in the
nucleon densities of \cite{Diehl:2005jf,Gockeler:2006zu}.
The result is  
much simpler but still contains a dipole term
$\propto s_\perp^i \epsilon^{ij}\ms b_\perp^j$, which leads to a 
dependence on the direction of $b_\perp$ for fixed $s_\perp$,
\begin{eqnarray}
 \rho^{n}(b_\perp,s_\perp)
 &=& \int_{-1}^{1} dx\, x^{n-1} \rho(x,b_\perp,s_\perp) 
\nonumber \\
\quad
&=& \frac{1}{2} \biggl[\ms A^\pi_{n0}(b_\perp^2) 
  - \frac{s_\perp^i \epsilon^{ij}\ms b_\perp^j}{m_\pi}\,
%    \frac{\partial}{\partial b_\perp^2}\, 
B_{Tn0}^{\pi \prime}(b_\perp^2)
  \,\biggr] \,,
\label{density}
\end{eqnarray}
where $B_{Tn0}^{\pi \prime}=\partial_{b_\perp^2}B_{Tn0}^\pi$.
The $b_\perp$ dependent vector and tensor generalized form factors
(GFFs) of the pion, $A^\pi_{n0}$ and $B^\pi_{Tn0}$, are moments of the
GPDs:
\begin{eqnarray}
\int_{-1}^1 dx\, x^{n-1} H^\pi(x,\ix0,b^2_\perp) 
&=& A^\pi_{n0}(b^2_\perp) \,,
\nonumber\\
\int_{-1}^1 dx\, x^{n-1} E_T^\pi(x,\ix0,b^2_\perp) 
&=& B^\pi_{T n0}(b^2_\perp) \,.
  \label{EqGPDmoments}
\end{eqnarray}
To this day, next to nothing is known about the signs and sizes of the
$B^\pi_{Tn0}$.  Since these GFFs determine the dipole-like distortion
of the quark density in the transverse plane, non-vanishing
$B^\pi_{Tn0}$ would imply a  
surprising \textsl{non-trivial
transverse spin structure of the pion}.  A computation of the
$B^\pi_{Tn0}$ from first principles in lattice QCD therefore provides
crucial insight into the pion structure.

Lattice QCD calculations give access to GFFs $F(t) =A^\pi_{n0}(t),
B^\pi_{Tn0}(t)$ in momentum space, which are related to the impact
parameter dependent GFFs $F(b_\perp^2) =A^\pi_{n0}(b_\perp^2),
B^\pi_{Tn0}(b_\perp^2)$ by a Fourier transformation
\begin{equation}
F(b_\perp^2) =
(2\pi)^{-2} \int d^2\Delta_\perp\,
  e^{-i b_\perp \cdot \Delta_\perp} F(t=-\Delta_\perp^2)\,,
  \label{Fourier}
\end{equation}
where $\Delta_\perp$ is the transverse momentum transfer.
The momentum-space GFFs $B^\pi_{Tn0}(t)$ parameterize pion matrix
elements of local tensor quark operators,
\begin{multline}
\langle \pi^+(P')| \mathcal{O}_T^{\mu\nu\mu_1\cdots\mu_{n-1}}
   | \pi^+(P)\rangle
= \mathcal{AS}\, \frac{\bar{P}^{\mu} \Delta^{\nu}
                     - \Delta^\mu \bar{P}^{\nu}}{m_\pi}
 \\
\times
  \sum^{n-1}_{\substack{i=0\\\textrm{even}}}
  \Delta^{\mu_1} \cdots \Delta^{\mu_i}\bar
  P^{\mu_{i+1}} \cdots \bar P^{\mu_{n-1}} B^\pi_{Tni}(t) \
    \label{tensorME}
\end{multline}
with $\bar P= \frac{1}{2} (P'+P)$, $\Delta = P'-P$ and $t=\Delta^2$.
Here $\mathcal{AS}$ denotes symmetrization in $\nu,\ldots,\mu_{n-1}$
followed by anti-symmetrization in $\mu,\nu$ and subtraction of traces
in all index pairs.  The tensor operators are given by
\begin{equation}
 \label{eq:gff:def}
  \mathcal{O}_T^{\mu\nu\mu_1\cdots\mu_{n-1}}
= \mathcal{AS}\; \overline{q}\, i\sigma^{\mu\nu}\ms
      i\Dlr^{\mu_1} \cdots\ms i\Dlr^{\mu_{n-1}}\ms q
\end{equation}
with $\Dlr = (\Dr - \Dl)/2$ and all fields taken at
space-time point $z=0$.  The analogous matrix elements of local vector quark
operators are parameterized by $A^\pi_{n0}(t)$ as specified in
\cite{Brommel:2005ee}.
For definiteness we consider in the following 
%the GFFs
$\smash{A^{\pi,u}_{n0}(t)}$ and $\smash{B^{\pi,u}_{T n0}(t)}$ for
up-quarks in a $\pi^+$.  Their counterparts for down-quarks and for
$\pi^-$ or $\pi^0$ 
readily follow from
isospin invariance \cite{Diehl:2005rn}, since Wilson fermions preserve
flavor symmetry.  
We note that $A^{\pi,u}_{10}(t)$ is
identical to the electromagnetic pion form factor $F_\pi(t)$, which we
investigated in detail in~\cite{Brommel:2006ww}.

\begin{figure}[t]
\vspace{-1mm}
\bc
\includegraphics[width=8.cm,angle=0,clip=true]
{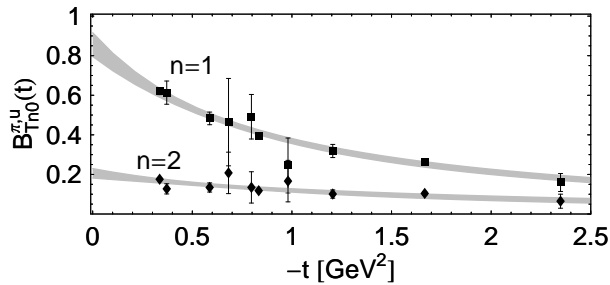}
\caption{\label{FigBbarT10} Lattice results at $\beta=5.29$ and
  $m_\pi\approx 600 \mev$ for the first two generalized form factors
  $B^{\pi,u}_{T n0}(t)$ for up-quarks in the $\pi^+$.  
The shaded bands show $p$-pole parameterizations.}
\ec \vspace{-5mm}
\end{figure}

{\em Lattice QCD results.}---
Based on our simulations with Wilson gluons
and dynamical, non-perturbatively ${\mathcal O}(a)$ improved Wilson
fermions with $n_f=2$, we have evaluated the matrix
elements in Eq.~(\ref{tensorME}) for $n=1,2$ and momentum transfers up
to $-t \approx 3 \gev^2$.
Configurations  
were generated at four
different couplings $\beta=5.20$, $5.25$, $5.29$, $5.40$ with up to
five different $\kappa=\kappa_{\mathrm {sea}}$ values per $\beta$, on
lattices of sizes $V\times T=16^3\times 32$ and $24^3\times 48$.  We have
set the lattice scale $a$ using a Sommer parameter of $r_0=0.467\fm$
\cite{Khan:2006de}.  The pion masses are as low as $400
\mev$, spatial volumes are as large as $(2.1 \fm)^3$, and lattice
spacings are below $0.1\fm$ (see \cite{Brommel:2006ww} for a list of lattice parameters).
The computationally demanding disconnected contributions present for
even $n$ are not included.  For the tensor GFFs $B^\pi_{Tn0}$ we
expect them to be small in the physical limit, since they require
a chirality flip on a quark line and are thus suppressed by the quark
mass \cite{Gockeler:2005cj}.   
All results 
were transformed to the $\overline{\mathrm{MS}}$ scheme at a scale of $4
\gev^2$ using non-perturbative renormalization \cite{reno}.  Further
information on the  
computation of GFFs in lattice QCD can be found,
e.g., in \cite{Gockeler:2003jf,Brommel:2006ww}, and details
of the present analysis will be given in \cite{Brommel:2007xx}.

\begin{figure}[t]
\vspace{-1mm}
\bc
\includegraphics[width=7.7cm,angle=0,clip=true]{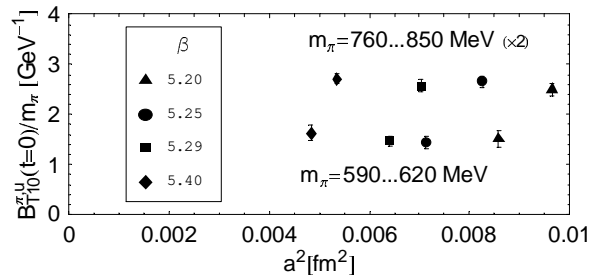}
\caption{\label{FigDiscrErr} Study of discretization errors in
$B^{\pi,u}_{T10}(\t0)/m_\pi$.}
\ec
\vspace{-5mm}
\end{figure}
As an example we show in Fig.~\ref{FigBbarT10} the $t$ dependence of
\smash{$B^{\pi,u}_{T (n=1,2)0}$} at $\beta=5.29$ and  
$m_\pi\approx 600 \mev$.  The extrapolation to the forward
limit $t=0$ requires a parameterization of the $t$ dependence of the
lattice results.  
As the statistics and $t$ range of our data is not yet 
sufficient for sophisticated multi-parameter fits, we use a standard $p$-pole form
$F(t)=F_0\ms /[\ms 1 - t/(p\ms m_p^2) \ms]^{\ms p}$,
where the forward value $F_0=F(\t0)$ and the $p$-pole mass $m_p$ are free parameters for each GFF. 
Good fits are obtained in a wide range of $p$, with a preference for relatively low values.  
On the other hand,
a regular behavior of $\rho^{n}(b_\perp,s_\perp)$
in the limit $b_\perp \to 0$ (which is of course inaccessible in a
lattice calculation) requires $p>3/2$ for $B^{\pi,u}_{Tn0}(t)$ \cite{Diehl:2005jf}.
We therefore take $p=1.6$ in the following.
For the examples in Fig.~\ref{FigBbarT10} we obtain
$B^{\pi,u}_{T10}(\t0)\!=\!0.856(60)$ with $m_p\!=\!0.949(57) \gev$, 
and $B^{\pi,u}_{T20}(\t0)\!=\!0.206(24)$ with $m_p\!=\!1.239(30)
\gev$. 
We stress that our final results show only a mild dependence on the chosen value of $p$.
Taking, e.g., $p\!=\!2$, which 
%corresponds to 
gives the power behavior for
$t\!\to\! -\infty$ expected from dimensional counting, changes our fits of
\smash{$B^{\pi,u}_{Tn0}$}  by less than the statistical errors 
even beyond the region $-t < 3 \gev^2$ where  
we have data \cite{Brommel:2007xx}.

Before discussing potential discretization and finite size effects as
well as the pion mass dependence of our results we note that, due to
the prefactor $m_\pi^{-1}$ in the parameterization~(\ref{tensorME}),
the GFFs $B^{\pi}_{Tn0}(t)$ must vanish like $m_\pi$ for $m_\pi \to 0$
\cite{Diehl:2005rn}.  This is also required to ensure that
the densities in Eq.~(\ref{density}) stay positive and finite in the chiral limit.
In the following we therefore consider the ratio $B^{\pi}_{Tn0}\ms
\big/m_\pi$, which tends to a constant at $m_\pi =0$.

Figure~\ref{FigDiscrErr} shows the dependence of $B^{\pi,u}_{T10}(\t0)
\big/m_\pi$ on the lattice spacing $a$ for two ranges of pion masses,
where we have excluded those lattice data points which are most
strongly affected by finite volume corrections (see below).  We conclude that
discretization errors are smaller than the statistical errors and
neglect any dependence of the GFFs on $a$ in the following analysis.

\begin{figure}[t]
\vspace{-1mm}
\bc
\includegraphics[width=7.5cm,angle=0]{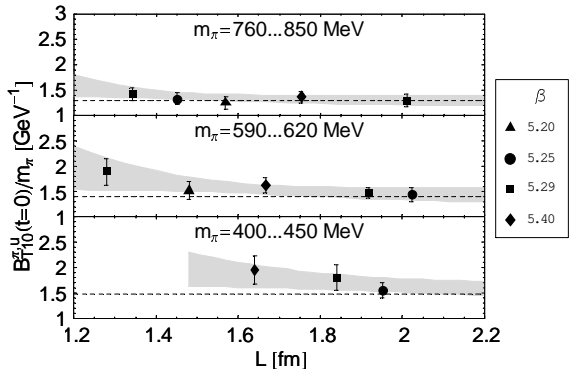}
\caption{\label{FigBT10finVol} Study of finite size effects 
of $B^{\pi,u}_{T10}(\t0)\big/m_\pi$.   
Shaded bands represent a combined fit
(restricted to $m_\pi L > 3$)
in $m_\pi$ and $L$ as described in the text.  
The dashed lines show the infinite-volume limit of the fit.}
\ec
\vspace{-6mm}
\end{figure}

Figure~\ref{FigBT10finVol} shows  
the volume dependence of 
$B^{\pi,u}_{T10}(\t0) \big/m_\pi$ for three
different ranges of $m_\pi$.  
The finite volume corrections to the
matrix elements with $n=1,2$ in Eq.~(\ref{tensorME}) are known to
leading order in chiral perturbation theory (ChPT) \cite{Manashov:2007qr}.
For $m_\pi L\gg 1$ the leading correction to $B^{\pi}_{Tn0}(\t0)
\big/m_\pi$ is proportional to $m_\pi^2 \exp(-m_\pi L)$ up to powers
of $(m_\pi L)^{-1/2}$, where $L$ is the spatial extent of the lattice.
Although our analysis includes pion masses as low as $400 \mev$, we
feel that a quantitative application of the chiral expansion requires
lattice computations at even lower values of $m_\pi$ and probably the
inclusion of higher-order terms.  We take however the result of
\cite{Manashov:2007qr} as a guide to estimate the $L$ dependence of
our lattice data, fitting $B^{\pi,u}_{T10}(\t0) \big/m_\pi$ to the
form $c_0 + c_1\ms m_\pi^2 + c_2\ms m_\pi^2 \exp(-m_\pi L)$.
This fit, represented by shaded bands in Fig.~\ref{FigBT10finVol},
gives $B^{\pi,u}_{T10}(\t0)=1.47(18)$ GeV$^{-1}$ at $L\!=\!\infty$
and $m_\pi\!\sim\!$ $440$~MeV, compared
to $B^{\pi,u}_{T10}(\t0)=1.95(27)$ GeV$^{-1}$ at 
$L\!\sim\! 1.65$ fm as represented by the diamond 
in the lowest panel of Fig.~\ref{FigBT10finVol}.
The typical corrections for $B^{\pi,u}_{T20}(\t0) \big/m_\pi$ are similar.
Within present statistics, we do not see a clear volume dependence of the
corresponding $p$-pole masses for $n=1,2$.

\begin{figure}[t]
\vspace{-1mm}
\bc
\includegraphics[width=7.2cm,angle=0]{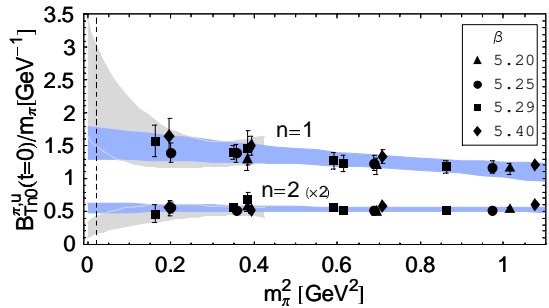}
\caption{\label{FigBbarT10mPi2} Pion mass dependence of
  $B^{\pi,u}_{Tn0}(\t0)\big/m_\pi$.  The shaded bands represent fits
  %linear in $m_\pi^2\protect\rule{0pt}{2.1ex}$
  as explained in the text.}
\ec 
\vspace{-4mm}
\end{figure}

The pion mass dependence of 
$B^{\pi,u}_{Tn0}(\t0)\big/m_\pi$ 
is shown in Fig.~\ref{FigBbarT10mPi2}.  
The darker shaded bands show fits based on the ansatz 
we just described. 
Data points and error bands have been
shifted to $L\!=\!\infty$.
For \mbox{$m_\pi= 140 \mev$} we obtain $B^{\pi,u}_{T10}(\t0)
\big/m_\pi = 1.54(24)\ms
\gev^{-1}$ with $m_p =0.756(95) \gev$, and
$B^{\pi,u}_{T20}(\t0) \big/m_\pi = 0.277(71)\ms \gev^{-1}$ with $m_p= 1.130(265) \gev$,
where in both cases we have set $p=1.6$.
The errors of the forward values include
the uncertainties from finite volume effects.
The light shaded bands in Fig.~\ref{FigBbarT10mPi2} show fits 
restricted to $m_\pi<650$ MeV using
1-loop ChPT \cite{Diehl:2005rn} plus the volume dependent term 
$c_2\ms m_\pi^2 \exp(-m_\pi L)$. 
We note that the ChPT-extrapolation gives larger values for 
$B^{\pi,u}_{T10}(\t0)$ at the physical point 
than the linear extrapolation in $m_\pi^2$.

\begin{figure}[t]
\vspace{+0mm}
\bc
\includegraphics[width=8.7cm,angle=0]{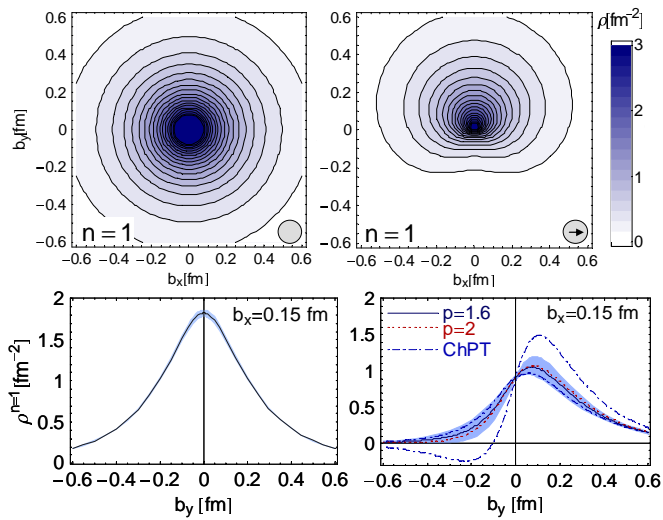}
\caption{\label{densities1} The lowest moment of the 
densities of unpolarized (left) and transversely polarized
(right) up-quarks in a $\pi^+$ together with corresponding
profile plots. The quark spin is
oriented in the transverse plane as indicated by the arrow.
The error bands in the profile plots show the uncertainties
in $B^{\pi,u}_{T10}(\t0)\big/m_\pi$ and the $p$-pole masses at 
$m_\pi^{\text{phys}}$ from a linear extrapolation.
The dashed-dotted lines show the uncertainty
from a ChPT-extrapolation (light shaded band in Fig.~5).} 
\ec
\vspace{-6mm}
\end{figure}

To compute the lowest two moments of the density in
Eq.~(\ref{density})  
we further need the GFFs $A^\pi_{n0}(t)$ with
$n=1,2$.  For $A^{\pi,u}_{10}(t) =F_\pi(t)$ we refer to our results in
\cite{Brommel:2006ww}.  A detailed analysis of $A^{\pi,u}_{20}(t)$
will be presented in \cite{Brommel:2007xx}, and first results 
are given in \cite{Brommel:2005ee}.  
We fit $A^{\pi,u}_{n0}(t)$ to a $p$-pole parameterization with $p=1$,
which provides an excellent description of the lattice data and is
consistent with power counting for $t\to -\infty$.
Fourier transforming the parameterizations of the momentum-space GFFs we
obtain the densities $\rho^{n}(b_\perp,s_\perp)$. 
In Fig.~\ref{densities1} we show 
$\rho^{n=1}(b_\perp,s_\perp)$
for up-quarks in a $\pi^+$ together with corresponding profile
plots for fixed $b_x$.
Compared to the unpolarized case on the left, 
the right-hand side of Fig.~\ref{densities1} shows 
strong distortions for
transversely polarized quarks and thus a pronounced spin structure.
The difference between $p=1.6$ and $p=2$ for
$B^{\pi,u}_{Tn0}$ is negligible within errors.
%%%%%%%%%%%%%%%%
The negative values of the density on the lower right in Fig.~\ref{densities1},
obtained for the maximal values of $B^{\pi,u}_{T10}$ from the chiral extrapolations 
in Fig.~\ref{FigBbarT10mPi2}, are unphysical. 
They show that 1-loop ChPT cannot be regarded as quantitatively reliable in this case
and provides only a rough idea of the uncertainties related to the chiral extrapolation.
%%%%%%%%%%%%%%%%
%%%%%%%%%%%%%%%%
From Eq.~\eqref{density} we obtain an average transverse
shift
\begin{align}
  \label{shift}
\langle b^y_\perp \rangle_{n}
 &= \frac{\int d^2 b_\perp\, b^{y}_\perp\ms \rho^{n}(b_\perp,s_\perp)}{%
          \int d^2 b_\perp\, \rho^{n}(b_\perp,s_\perp)}
= \frac{1}{2 m_\pi}\, \frac{B^{\pi}_{T n0}(\t0)}{A^{\pi}_{n0}(\t0)}
\end{align}
in the $y$ direction for a transverse quark spin $s_\perp =(1,0)$ in the $x$ direction.
Our lattice results give $\langle b^y_\perp \rangle_{1} = 0.151(24)\fm$ 
and $\langle b^y_\perp \rangle_{2} = 0.106(28) \fm$.

Let us compare our results for $B^{\pi}_{Tn0}$ with those for the
analogous GFFs $\overline B_{Tn0}$ that describe the dipole-like
distortion in the density of transversely polarized quarks in an
unpolarized nucleon.  The corresponding average transverse shift is
$\langle b^y_\perp \rangle_{n} = \overline B_{Tn0}(\t0) \big/\bigl( 2 m_N
A_{n0}(\t0) \bigr)$, where $A_{n0}(\t0)$ is the $n$-th moment of the
unpolarized quark distribution.  With the lattice results of
\cite{Gockeler:2006zu} we find $\langle b^y_\perp \rangle_{1} = 0.154(6) \fm$ 
and $\langle b^y_\perp \rangle_{2} = 0.101(8) \fm$ for up-quarks in the proton.
Remarkably, the distortion in the distribution of a transversely
polarized up-quark is within errors of the same strength in a $\pi^+$ and in the
proton.  An explanation of this finding  has recently been proposed in
the framework of quark models \cite{Burkardt:2007xm}.
 
The moments of the GPDs $E^\pi_T$ in the pion
and $\overline E_T$ in the nucleon can be connected with the
respective Boer-Mulders functions, which describe the correlation
between transverse spin and intrinsic transverse momentum of quarks in an
unpolarized hadron \cite{Boer:1997nt}.  They lead, e.g., to azimuthal
asymmetries in semi-inclusive deep inelastic scattering (SIDIS) and in
Drell-Yan lepton pair production.  
The density of quarks with transverse momentum $k_\perp$ and transverse spin $s_\perp$ in a
$\pi^+$ is given by
\begin{equation}
  \label{Boer-Mulders}
f(x,k_\perp,s_\perp) = \frac{1}{2} \biggl[ f_1^{\pi}(x,k_\perp^2)
  + \frac{s_\perp^i \epsilon^{ij}\ms k_\perp^j}{m_\pi}\,
    h_1^{\pi \perp}(x,k_\perp^2) \biggr]
\end{equation}
in terms of the unpolarized distribution $f_1^{\pi}$ and the Boer-Mulders
function $h_1^{\pi \perp}$. 
We notice the close similarity between~\eqref{Boer-Mulders} and the impact parameter
density~\eqref{density}, but emphasize that $k_\perp$ and $b_\perp$
are \textsl{not} Fourier conjugate variables.  A dynamical relation
between $k_\perp$ and $b_\perp$ dependent densities 
was proposed in \cite{Burkardt:2005hp} and implies
$\smash{h_{1}^{\perp,\pi}} \sim -E^{\pi\phantom{,}}_T$ for the
distribution appearing in SIDIS---we recall that $h_1^{\perp\pi}$ is
time reversal odd and thus enters with opposite signs in SIDIS and Drell-Yan production
\cite{Collins:2002kn}.  With this relation, our results for
$B^{\pi}_{T n0}$ imply that the Boer-Mulders function
for up-quarks in a $\pi^+$ is large and negative, 
and that its ratio to the unpolarized distribution
is similar for up-quarks in a $\pi^+$ and in a proton.

{\em Conclusions.}---
We have calculated the first two moments of the quark tensor GPD
$E_{T}^{\pi}$ in the pion.  We find that the spatial distribution of
quarks is strongly distorted if they are transversely polarized, 
revealing a non-trivial spin structure of the pion.  The effect has the
same sign and very similar magnitude as the corresponding distortion
in the nucleon \cite{Gockeler:2006zu}.  Assuming the
relation between impact parameter and transverse momentum densities
proposed in \cite{Burkardt:2005hp} this suggests that
all Boer-Mulders functions for valence quarks may be alike, as argued
in \cite{Burkardt:2007xm}.
The large size of the effect might give new insight into the mechanism
responsible for the large $\cos(2\phi)$ azimuthal asymmetry observed
in unpolarized $\pi p$ Drell-Yan production, which is sensitive to the
product $h_{1}^{\perp \pi} h_{1}^{\perp}$
\cite{Boer:1999mm}.  It
motivates future studies of azimuthal
asymmetries in unpolarized $\pi p$ and polarized $\pi p^{\uparrow}$
Drell-Yan production at COMPASS, the latter giving rise to a
$\sin(\phi+\phi_S)$ asymmetry sensitive to $h_{1}^{\perp \pi}
h_{1}^{}$, where $h_{1}^{}$ is the quark transversity distribution in
the nucleon \cite{Sissakian:2005yp}.
\begin{acknowledgments}
The numerical calculations have been performed on the Hitachi SR8000
at LRZ (Munich), apeNEXT and APEmille at NIC/DESY (Zeuthen) and BlueGene/Ls
at NIC/FZJ (J\"ulich), EPCC (Edinburgh) and KEK (by the Kanazawa group
as part of the DIK research program).  This work was supported by
DFG (Forschergruppe Gitter-Hadronen-Ph\"anomenologie and Emmy-Noether
program), by HGF (contract No.\ VH-NG-004) and by EU I3HP (contract
No.\ RII3-CT-2004-506078).
\vspace{-1mm}
\end{acknowledgments}
\vspace{-4mm}

%%%%%%%%%%%%%%%%%%%%%%%%%%%%%%%%%%%%%%%%%%%%%%%%%%%%%%%%%%%%%%%%%

\end{document}